\documentclass{article}
\usepackage{spconf,amsmath,graphicx}
\usepackage[pagebackref,breaklinks,colorlinks]{hyperref}
\usepackage{multirow}
\usepackage{textpos}
\usepackage[moderate]{savetrees}

\ninept

\title{Metric Learning for User-defined Keyword Spotting}

\name{
    \begin{tabular}{c}
    Jaemin Jung$^{1}$*, Youkyum Kim$^{1}$*, Jihwan Park$^{2, 3}$, Youshin Lim$^{2, 3}$, Byeong-Yeol Kim$^{2, 3}$, \\ 
    Youngjoon Jang$^{1}$, Joon Son Chung$^{1}$
    \end{tabular}
     \thanks{\hspace{-10pt}* These authors contributed equally.}}
\address{$^1$Korea Advanced Institute of Science and Technology, Daejeon, Republic of Korea \\       
          $^2$Hyundai Motor Company, $^3$42dot Inc., Seoul, Republic of Korea}
         
\usepackage{ctable}
\usepackage{extra_styles}
\usepackage[capitalize]{cleveref}
\usepackage{subcaption}
\usepackage{caption}

\crefname{section}{Sec.}{Secs.}  
\Crefname{section}{Section}{Sections}
\Crefname{table}{Table}{Tables}
\crefname{table}{Tab.}{Tabs.}
\crefname{algorithm}{Alg.}{Algs.}  
\Crefname{algorithm}{Algorithm}{Algorithms}

\begin{document}

\maketitle

\begin{abstract}
The goal of this work is to detect new spoken terms defined by users. While most previous works address Keyword Spotting (KWS) as a closed-set classification problem, this limits their transferability to unseen terms. The ability to define custom keywords has advantages in terms of user experience.

In this paper, we propose a metric learning-based training strategy for user-defined keyword spotting. In particular, we make the following contributions: 
(1) we construct a large-scale keyword dataset with an existing speech corpus and propose a 
filtering method to remove data that degrade model training;
(2) we propose a metric learning-based two-stage training strategy, and demonstrate that the proposed method improves the performance on the user-defined keyword spotting task by enriching their representations;
(3) to facilitate the fair comparison in the user-defined KWS field, we propose unified evaluation protocol and metrics.

Our proposed system does not require an incremental training on the user-defined keywords, and outperforms previous works by a significant margin on the Google Speech Commands dataset using the proposed as well as the existing metrics.
\end{abstract}
\begin{textblock*}{.8\textwidth}[.5,0](0.5\textwidth, -.455\textwidth)
\centering
{\url{https://mm.kaist.ac.kr/projects/kws}}
\end{textblock*}
\section{Introduction}
\label{sec:intro}

As an entry point to many speech-enabled systems, the performance of KWS systems is critical to providing a satisfactory user experience. While most existing KWS systems are based on a set of pre-defined keywords, the ability to define custom keywords can significantly improve user experience. For instance, the use of different keywords on different devices prevents accidental wake of nearby devices, and it also provides a layer of security by preventing access from strangers.

Most recent KWS methods~\cite{choi2019temporal, sorensen2020depthwise, kim2021broadcasted, berg2021keyword, huh2021metric} are based on classification networks that distinguish between target keywords and non-target noises. On the other hand, keyword spotting is  closer to a detection task than a classification task, where the keywords are spotted from a range of unknown sounds. 
Therefore, user-defined keyword spotting can be naturally formulated as a metric learning problem, similar to face and speaker verification.
To this end, we propose a metric learning-based training strategy to learn effective representations that can be used in query-by-keyword systems.

In many metric learning applications such as face~\cite{schroff2015facenet, sohn2016improved, cao2018vggface2} and speaker recognition~\cite{chung2018voxceleb2, wang2019centroid, kwon2021ins}, it has been demonstrated that the number of training classes strongly correlates with performance on the downstream task. 
However, the popular Google Speech Commands (GSC) dataset~\cite{warden2018speech} only contains 35 classes, which is insufficient to facilitate good generalisation.
To generate additional training data, previous works~\cite{awasthi2021teaching, mazumder2021few, huang2021query} extract keywords from Automatic Speech Recognition (ASR) datasets with a forced aligner~\cite{mcauliffe2017montreal} and then use them as training data. However, this approach does not guarantee the quality of the extracted keyword data. To mitigate this issue, we propose a new filtering method based on Character Error Rate (CER) to verify whether or not the extracted keywords are properly segmented using a pre-trained speech recognition model. 

We conduct a wide variety of experiments using a range of objective functions borrowed from recent few-shot learning literature. 
Moreover, we propose a two-stage training strategy where we first pre-train the model on a large-scale out-of-domain corpus, then fine-tune the model on smaller in-domain data.
We perform extensive ablations on the amount of training data, the two-stage training strategy, and various parameters to find their effects on KWS performance.

Finally, we list a number of metrics that are suitable for the user-defined keyword spotting as a detection task. While most existing works use the classification accuracy to evaluate their systems, the metrics of interest to the developers of KWS applications are the False Alarm Rates (FAR) and False Rejection Rates (FRR) at given operating points, represented on a Detection Error Tradeoff (DET) curve. Moreover, there is no standard evaluation protocol in the field that enables fair comparisons with different works. To this end, we propose an evaluation protocol that is relevant to the detection task. We evaluate the performance of the proposed methods on GSC dataset, using the existing metrics. Our model outperforms comparable works by a significant margin.


\subsection{Related Works}
\label{sec:related}

In recent years, with the advances in deep learning technology, deep neural networks have been applied to the keyword spotting research. \cite{sainath2015convolutional} utilises Convolutional Neural Network (CNN) to KWS, and \cite{zhang2017hello} explores depth-wise separable convolution and point-wise convolution for the KWS task. 

With the intimate involvement of technology into our daily lives, personalized services and privacy have become more important. As a result, there has been growing attention on {\em user-defined} keyword spotting technology. The user-defined KWS task can be viewed from two perspectives -- as a \emph{classification task} or a \emph{detection task}.

\cite{awasthi2021teaching, mazumder2021few, liu2021keyword} solve user-defined keyword spotting task as a classification task. \cite{liu2021keyword} classifies non-target keywords into multiple classes in pre-training, and re-trains the model on the target keywords with data augmentation. \cite{awasthi2021teaching} replaces the last linear layer with a randomly initialized linear layer during fine-tuning. \cite{mazumder2021few} reinforces the model's representation capability by pre-training their model on the multilingual keyword dataset. 

Some works~\cite{huang2021query, parnami2022few} approach the problem of KWS as a detection task. \cite{huang2021query} designs a model architecture with multi-head attention layers and introduces soft-triple loss, which is a combination of triplet loss and softmax loss for learning feature representations. \cite{parnami2022few} proposes metric learning-based prototypical network that can effectively extract distinctive features to detect user-defined keywords. The method still requires an additional incremental training process to adapt the model to the target user-defined keywords. 
In contrast, our method does not require any incremental training during enrollment.

While there have been significant advances in keyword spotting algorithms, the field still suffers from the lack of diverse training data.
To overcome this shortage problem, few recent studies try to extract keyword data from a large-scale speech corpus using forced aligners~\cite{mcauliffe2017montreal}. For example, \cite{awasthi2021teaching, huang2021query} and~\cite{mazumder2021few} extract keyword data from the LibriSpeech dataset~\cite{panayotov2015librispeech} and the Common Voice dataset~\cite{ardila2019common} respectively.
\cite{awasthi2021teaching, mazumder2021few, huang2021query} all use force-align transcripts to construct KWS datasets without verifying the aligned results.
They also divide the datasets into training and test splits without considering the duplication of keywords between the sets.

In contrast, to validate the aligned keyword data, we evaluate CER on each of the keyword instances with a pre-trained speech recognition model and decide whether to include each data into the final training set.
Moreover, we ensure that the keywords used in the pre-training or fine-tuning do not appear in the user-defined test data.
Therefore, our experimental setup better addresses the {\em user-defined} keyword spotting problem in comparison to the previous literature.
\section{Methodology}
\label{sec:method}


\subsection{Large-scale Keyword Dataset}\label{protonet}
We construct a new large-scale keyword dataset, named LibriSpeech Keywords (LSK), consisting of 1,000 keyword classes extracted from the LibriSpeech corpus~\cite{panayotov2015librispeech}. We utilise a pre-trained wav2vec 2.0 model~\cite{baevski2020wav2vec} to force-align individual words from utterance-level labels. The wav2vec 2.0 model is pre-trained on 960 hours of unlabeled audio from LibriSpeech dataset and fine-tuned on the same audio with the corresponding transcript.
The extracted keywords are truncated by 1 second to include noises or utterances that may occur before or after the keyword in a real-world scenario.

Unlike previous works~\cite{mazumder2021few, huang2021query} which simply use the outputs of the forced aligner as their training dataset, we also validate the quality of the collected data. First, we compute CER score on each keyword in our dataset with the pre-trained wav2vec 2.0 model to filter misaligned examples which should not be used during the training step. Second, the 13 most frequent words and one-letter words are removed, because they consist mostly of articles and prepositions which are hard to recognise. Finally, 10 keywords in GSC dataset that are used as the user-defined keywords are removed. From this filtering process, we select the 1,000 most frequent keywords as our training data, followed by randomly sampling 1,000 instances per keyword. Note that our LSK dataset is only used in the {\em pre-training} stage.

\begin{figure}[t!]
  \centering
  \includegraphics[width=0.85\linewidth]{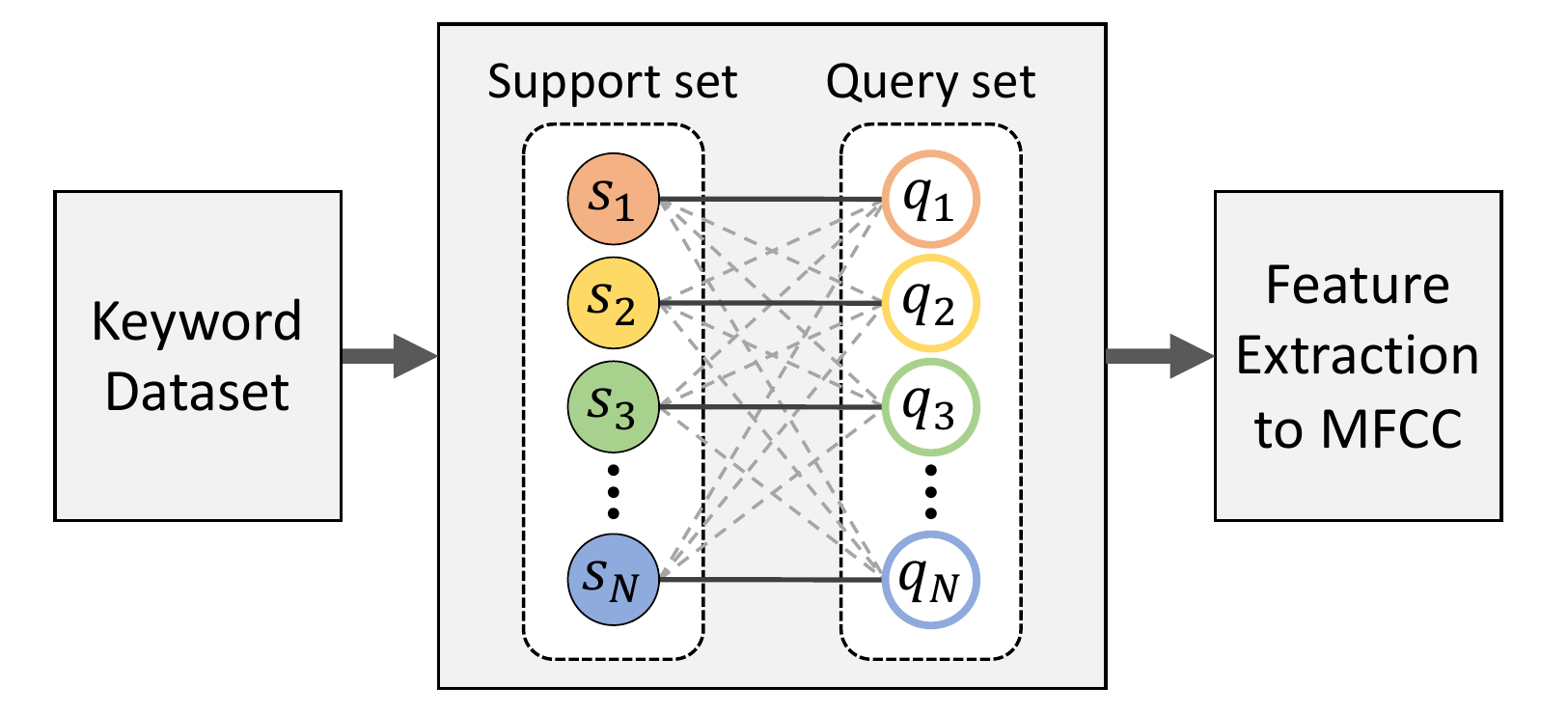}
  \vspace{-4mm}
  \caption{Batch configuration for metric learning-based training. $s_i$ and $q_i$ stand for samples from $i$-th keyword class in the support set and the query set, respectively.
  $N$ denotes size of a mini-batch. Solid lines connect positive pairs, while dotted lines represent negative pairs.}
  \label{fig:overall_framework}
  \vspace{-4mm}
\end{figure}

\subsection{Training Strategy}
Our training strategy is divided into pre-training and fine-tuning stages. In the pre-training stage, our model is trained on the proposed LSK dataset containing out-of-domain data to represent spoken words in the discriminative embedding space. In the fine-tuning stage, our model is fine-tuned using only 25 keywords of the in-domain GSC dataset. During inference, we consider the remaining 10 keywords of GSC dataset to be user-defined keywords, simulating the real-world deployment scenario. 
Note that the GSC and LSK datasets exhibit different characteristics in terms of acoustics and word isolation, hence the need for fine-tuning.

\subsection{Objective Functions}

In this paper, we compare the baseline softmax loss and two metric learning-based objective functions. $N$ denotes the number of utterances per mini-batch.

\newpara{Softmax.} 
Softmax loss consists of a softmax function followed by a multi-class cross-entropy loss. 
It is formulated as:
\begin{equation}
\footnotesize
L_\text{S}=-\frac{1}{N}\sum_{i=1}^N\log \frac{e^{\mathbf{W}^T_{y_i}\mathbf{x}_i+b_{y_i}}}{\sum_{j=1}^C e^{\mathbf{W}^T_{j}\mathbf{x}_i+b_{j}}},
\end{equation}  
where $\mathbf{W}$ and $b$ are learnable parameters, $\mathbf{x}_i$ is the feature vector, $\mathbf{y}_i$ is the class label for corresponding $\mathbf{x}_i$. Here, $C$ is the number of keyword classes.  
The loss function does not explicitly enforce intra-class compactness and inter-class separation. We use the softmax loss as the baseline objective function.

\newpara{AM-Softmax.}
Additive Margin Softmax (AM-Softmax) loss~\cite{wang2018additive, wang2018cosface} introduces a margin to the original Softmax Loss. First, the weights and the input vectors are normalised in the softmax loss such that the posterior probability only relies on cosine of the angle between the weights and the input vectors:

\begin{equation}
\footnotesize
L_\text{N}=-\frac{1}{N}\sum_{i=1}^N\log \frac{e^{{\cos(\theta_{{y_i}, i}})}}{\sum_{j} e^{{\cos(\theta_{{j}, i}})}},
\end{equation}
where $\cos{(\theta_{j,i})}$ is a dot product of normalised vector $\mathbf{W}_j$ and $\mathbf{x}_i$.

These embeddings are still not sufficiently discriminative because the equation only penalises the classification error. Cosine margin $m$ is incorporated into the equation to mitigate this issue:
\begin{equation}
\footnotesize
L_\text{C}=-\frac{1}{N}\sum_{i=1}^N\log \frac{e^{s({\cos(\theta_{{y_i}, i})-m})}}{e^{s({\cos(\theta_{{y_i}, i})-m)}} + \sum_{j\neq{y_i}} e^{s({\cos(\theta_{{j}, i})})}},
\end{equation}
where $s$ is a fixed scale factor to prevent gradient from getting too small in training phase.

\newpara{Angular Prototypical.}
The objective of prototypical loss~\cite{snell2017prototypical} is to learn effective representations by explicitly optimising the distance between the query and the prototype (support set).
In particular,
we use the Angular Prototypical (AP) loss~\cite{chung2020defence}, which replaces the Squared Euclidean distance metric in the regular prototypical loss function with the cosine distance. 
For each keyword in a mini-batch, we consider one utterance out of $M$ utterances as the query set, and the others as the support set.
We will assume that the query is $M$-th utterance from every keyword for simplicity.
Then the prototype (or centroid) for class $k$ is:
\begin{equation} \label{eqn:centroid}
\footnotesize
\textbf{c}_{k} = \frac{1}{M-1}\sum_{i = 1}^{M-1}{\textbf{e}_{k,i}},
\end{equation}

where $\textbf{e}_{k,i}$ denotes an embedding feature. We use a cosine-based similarity metric with learnable scale $w$ and bias $b$, as in the GE2E loss~\cite{wan2018generalized}.
\begin{equation} \label{eqn:cossim}
\footnotesize
\textbf{S}_{j,k} = w \cdot \cos(\textbf{e}_{j,M}, \textbf{c}_{k}) + b
\end{equation}

During training, each query example is
classified against $B$ classes in the mini-batch based on the softmax over distances to each keyword prototype:
\begin{equation} \label{eqn:protoloss}
\footnotesize
L_\text{AP} = -\frac{1}{B}\sum_{j=1}^{B}{\log\frac{e^{\textbf{S}_{j,j}}}{\sum_{k=1}^{B}{e^{\textbf{S}_{j,k}}}}}.
\end{equation}

\subsection{Batch Configuration}
In each mini-batch, only one pair is positive and the rest are all considered negative pairs. As shown in~\Fref{fig:overall_framework}, a positive pair consists of the same keyword but different audio data, whereas all the negative pairs consist of different keywords. For the prototypical-based networks, at least 2 samples per class per mini-batch are required.

\section{Experiments}
\label{sec:experiment}

\subsection{Datasets}

\newpara{Google Speech Commands (GSC).}
To simulate a real-world deployment scenario, we select GSC dataset, which contains 35 different keywords, as our target domain keyword set. As shown in~\Tref{tab:data_split}, we divide the GSC dataset into pre-defined, unknown, and user-defined splits. Note that only the pre-defined and unknown classes are used during the fine-tuning stage and we consider all classes in unknown split as one class.

\begin{table}[h!]
\centering
\resizebox{0.65\linewidth}{!}{
\begin{tabular}{|c|c|c|} 
\hline
{\textbf{Datasets}} & {\textbf{\# Classes}}& {\textbf{Keywords}} \\ 
\hline
Pre-defined & 10  & \begin{tabular}[c]{@{}c@{}}`Yes’, `No’, `Up’, \\ `Down’, `Left’, `Right’, `On’, \\`Off’, `Stop’, `Go'\end{tabular} \\ 
\hline
Unknown &15& \begin{tabular}[c]{@{}c@{}}`Bed’, `Bird’, `Cat’, \\`Dog’, `Wow’, `House’, \\ `Learn’, `Sheila’, `Tree’,\\`Happy’, `Marvin’, \\ `Backward’, `Follow’, \\`Forward’, `Visual’ \end{tabular} 
\\ 
\hline
User-defined &10& \begin{tabular}[c]{@{}c@{}}`Zero’, `One’, `Two’,`Three’, \\ `Four’,`Five’,`Six’,`Seven’, \\`Eight’, `Nine’\end{tabular} \\
\hline
\end{tabular}}
\vspace{-3mm}
\caption{Google Speech Commands dataset splits for user-defined keyword spotting.}
\label{tab:data_split}
\vspace{-4mm}
\end{table}


\newpara{LibriSpeech.}
LibriSpeech~\cite{panayotov2015librispeech} dataset is a widely used speech corpus which contains 1,000-hour spoken sentences along with the corresponding transcripts. We construct our LibriSpeech Keywords (LSK) dataset using all of the training data in LibriSpeech.

\newpara{Korean Speech Commands.}
Korean Speech Commands~\cite{koreacommand} contains a 4,000-hour Korean speech set. We construct our Korean Speech Keywords (KSK) dataset following the same pipeline used for the LSK dataset.

\begin{table*}[t]
\centering
\resizebox{0.90\linewidth}{!}{
    \begin{tabular}{c c|ccc|ccc|ccc|ccc|ccc} 
    \specialrule{.1em}{.05em}{.05em}
    \multicolumn{2}{c|}{Training loss} & \multicolumn{3}{c|}{\quad\quad EER~$\downarrow$ \quad\quad} & \multicolumn{3}{c|}{\quad\quad Acc~$\uparrow$ \quad\quad} & \multicolumn{3}{c|}{\quad F1-score~$\uparrow$ \quad} & 
    \multicolumn{3}{c|}{FRR@FAR=2.5~$\downarrow$} & \multicolumn{3}{c}{FRR@FAR=10~$\downarrow$}  
    \\ 
    \hline
    Pre-train & Fine-tune & 1-shot & 5-shot & 10-shot & 1-shot & 5-shot & 10-shot & 1-shot & 5-shot & 10-shot & 1-shot & 5-shot & 10-shot & 1-shot & 5-shot & 10-shot                             
    \\ 
    \specialrule{.1em}{.05em}{.05em}
    \multicolumn{2}{c|}{\cite{liu2021keyword} w/ Inc. Training} & - & - & 9.0$\dag$ & - & - & - & - & - & - & - & - & 17.0$\dag$ & - & - & 8.3$\dag$      \\
    \hline
    \multirow{3}{*}{-}  & Softmax   & 17.31 & 9.52 & 7.79 & 69.57 & 84.13 & 84.10 & 0.68 & 0.84 & 0.84 & 44.47 & 24.30 & 19.83 & 24.17 & 8.83 & 6.03                 \\
                        & AM-Soft   & 17.43 & 8.91 & 7.20 & 63.43 & 84.60 & 86.97 & 0.63 & 0.85 & 0.87 & 55.10 & 21.33 & 18.30 & 26.57 & 7.73 & 5.10                 \\
                        & AP        & 20.47 & 9.33 & 8.50 & 61.37 & 80.30 & 80.13 & 0.60 & 0.80 & 0.80 & 56.53 & 26.60 & 23.00 & 35.47 & 8.57 & 6.93                 \\ 
    \hline
    \multirow{4}{*}{Softmax} & -       & 30.77 & 20.64 & 19.01 & 47.07 & 62.23 & 67.23 & 0.47 & 0.63 & 0.68 & 66.10 & 59.57 & 44.10 & 51.20 & 36.87 & 27.77                   \\
                             & Softmax & 16.91 & 11.00 & 9.20  & 69.47 & 83.23 & 85.47 & 0.68 & 0.83 & 0.85 & 48.33 & 26.67 & 21.67 & 25.10 & 11.67 & 8.67                     \\
                             & AM-Soft & 10.47 & 4.75 & 4.01   & 85.43 & 94.80 & 95.33 & 0.85 & 0.95 & 0.95 & 24.20 & 6.90  & 5.97  & 10.87 & 3.07 & 2.03             \\
                             & AP      & 10.10 & 5.20 & 3.77 & 83.53 & 94.23 & 95.00 & 0.83 & 0.94 & 0.95 & 23.00 & 7.67 & 5.47  & 10.23 & 3.40 & 2.20 \\ 
    \hline
    \multirow{4}{*}{AM-Soft} & -        & 34.78 & 26.87 & 22.65 & 41.73 & 56.83 & 63.30 & 0.43 & 0.58 & 0.64 & 75.77 & 75.00 & 61.53 & 61.80 & 51.23 & 38.87                     \\
                             & Softmax  & 23.60 & 15.38 & 13.80 & 53.17 & 70.50 & 77.93 & 0.54 & 0.70 & 0.78 & 65.23 & 44.07 & 37.87 & 41.73 & 22.73 & 18.27                      \\
                             & AM-Soft  & 10.88 & 6.54 & 5.64   & 85.13 & 92.57 & 93.63 & 0.85 & 0.93 & 0.94 & 26.40 & 11.87 & 9.60  & 11.63 & 4.80 & 3.50           \\
                             & AP       & 11.80 & 6.57 & 4.80 & 80.27 & 92.40 & 93.07 & 0.79 & 0.92 & 0.93 & 28.20 & 12.43 & 7.63 & 13.70 & 4.70 & 3.13 \\ 
    \hline
    \multirow{4}{*}{AP}      & -        & 32.70 & 24.81 & 21.01 & 41.23 & 60.03 & 69.37 & 0.45 & 0.61 & 0.70 & 80.50 & 74.03 & 57.67 & 59.60 & 50.77 & 36.23                         \\ 
                             & Softmax  & 15.81 & 10.97 & 8.87  & 70.07 & 80.77 & 83.33 & 0.71 & 0.81 & 0.84 & 55.10 & 29.60 & 20.77 & 25.83 & 12.07 & 7.57                                \\
                             & AM-Soft  & 8.08 & 5.31 & 4.27   & 88.53 & {\bf 94.03} & 95.53 & 0.88 & {\bf 0.94} & {\bf 0.96} & 17.10 & 7.50 & 5.67   & 6.90 & 3.37 & 2.60                             \\
                             & AP       & {\bf 7.77} & {\bf 4.49} & {\bf 3.24} & {\bf 89.97} & 93.93 & {\bf 95.97} & {\bf 0.90} & {\bf 0.94} & {\bf 0.96} & {\bf 16.77} & {\bf 6.67} & {\bf 4.20} & {\bf 5.93} & {\bf 2.47} & {\bf 1.20}            
                             \\
    \specialrule{.1em}{.05em}{.05em}
    \end{tabular}}
\vspace{-3mm}
\caption{Experimental results using different loss functions. All numbers are in percent (\%) except the F1-score.
{\bf FRR@FAR=10}: False Rejection Rate at False Alarm Rate of 10\%;
{\bf AM-Soft}: Additive Margin Softmax loss; 
{\bf AP}: Angular Prototypical loss;
{\bf $\dag$}: digitized from the DET curve on Figure 9(a) of \cite{liu2021keyword}.}
\label{tab:class_vs_metric}
\vspace{-1mm}
\end{table*}




\begin{table*}[t]
\centering
\resizebox{0.65\linewidth}{!}{
\begin{tabular}{c c c |ccccc} 
\specialrule{.1em}{.05em}{.05em}
Dataset              & \# Classes & \# Samples & {\quad EER~$\downarrow$\quad} & {\quad Acc~$\uparrow$\quad} & F1-score~$\uparrow$ & FRR@FAR=2.5~$\downarrow$ & FRR@FAR=10~$\downarrow$  \\ 
\hline
\multirow{4}{*}{LSK}    & 500   & 500       & 4.13      & 94.50     & 0.95           & 6.07              & 2.13              \\           
                        & 500   & 1,000     & 3.94      & 94.60     & 0.95           & 5.23              & 1.97              \\
                        & 1,000  & 500      & 3.63      & 95.07     & 0.95           & 5.13              & 1.47            \\
                        & 1,000  & 1,000    & 3.24      & {\bf 95.97}     & {\bf 0.96}     & 4.20              & {\bf 1.20}             \\ 
\hline
LSK+KSK              & 2,000  & 1,000   & {\bf 3.07}     & 95.63    & {\bf 0.96}          & {\bf 3.73}    & {\bf1.20}             \\
\specialrule{.1em}{.05em}{.05em}
\end{tabular}}
\vspace{-3mm}
\caption{The effect of pre-training data on final system performance. All numbers are in percent (\%) except for F1-score. All experiments in this table utilise the Angular Prototypical (AP) loss.}
\label{tab:samples}
\vspace{-4mm}
\end{table*}

\begin{figure*}[!t]
\centering
\vspace{-1mm}
\resizebox{\linewidth}{!}
{
     \begin{subfigure}[b]{0.3\textwidth}
         \centering
         \includegraphics[width=0.65\textwidth]{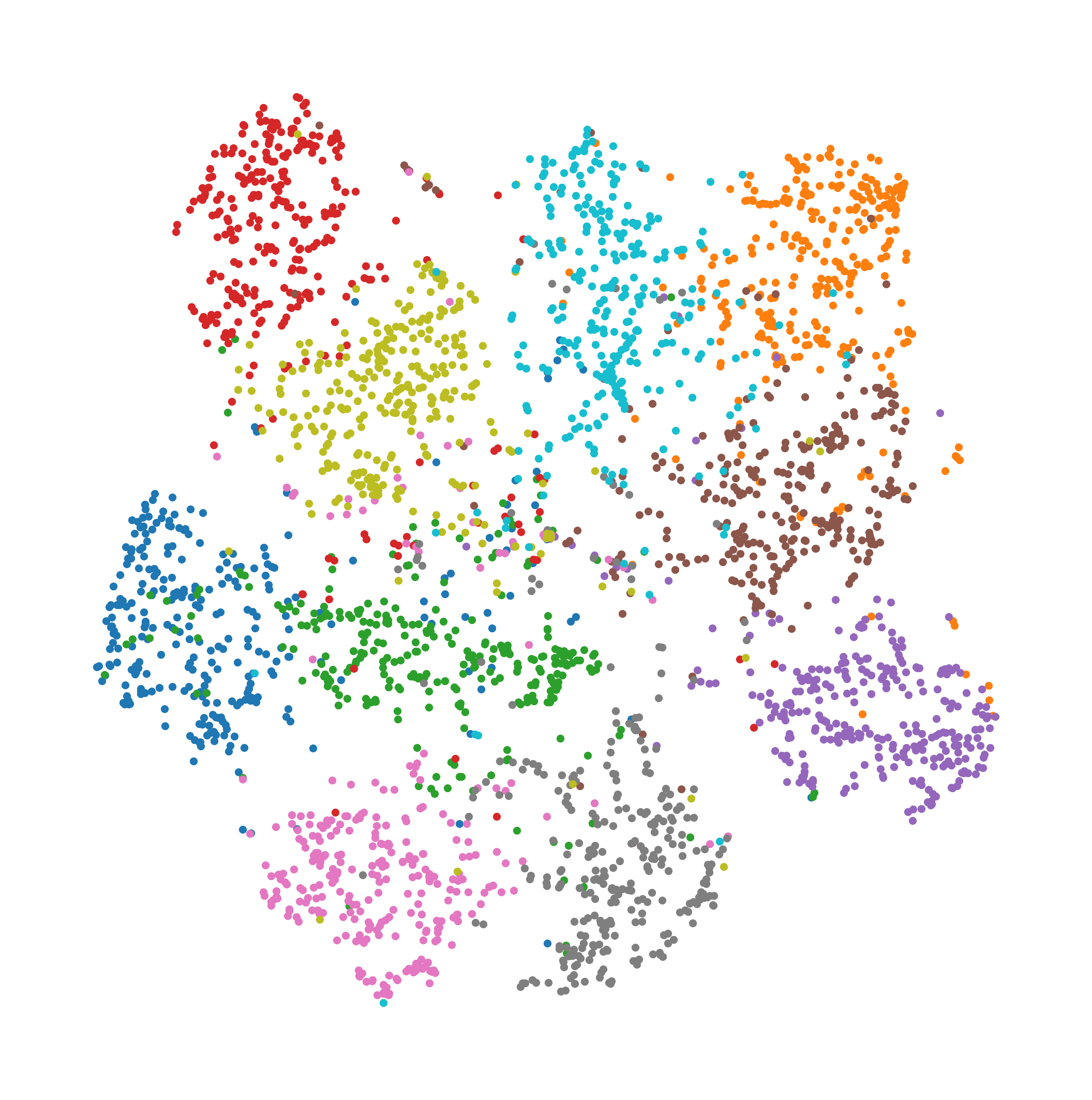}
         \caption{Trained only on the GSC dataset.}
         \hspace{1.0cm}
         \vspace{-2mm}
         \label{fig:tsne_GSC}
     \end{subfigure}
     \begin{subfigure}[b]{0.3\textwidth}
         \centering
         \includegraphics[width=0.65\textwidth]{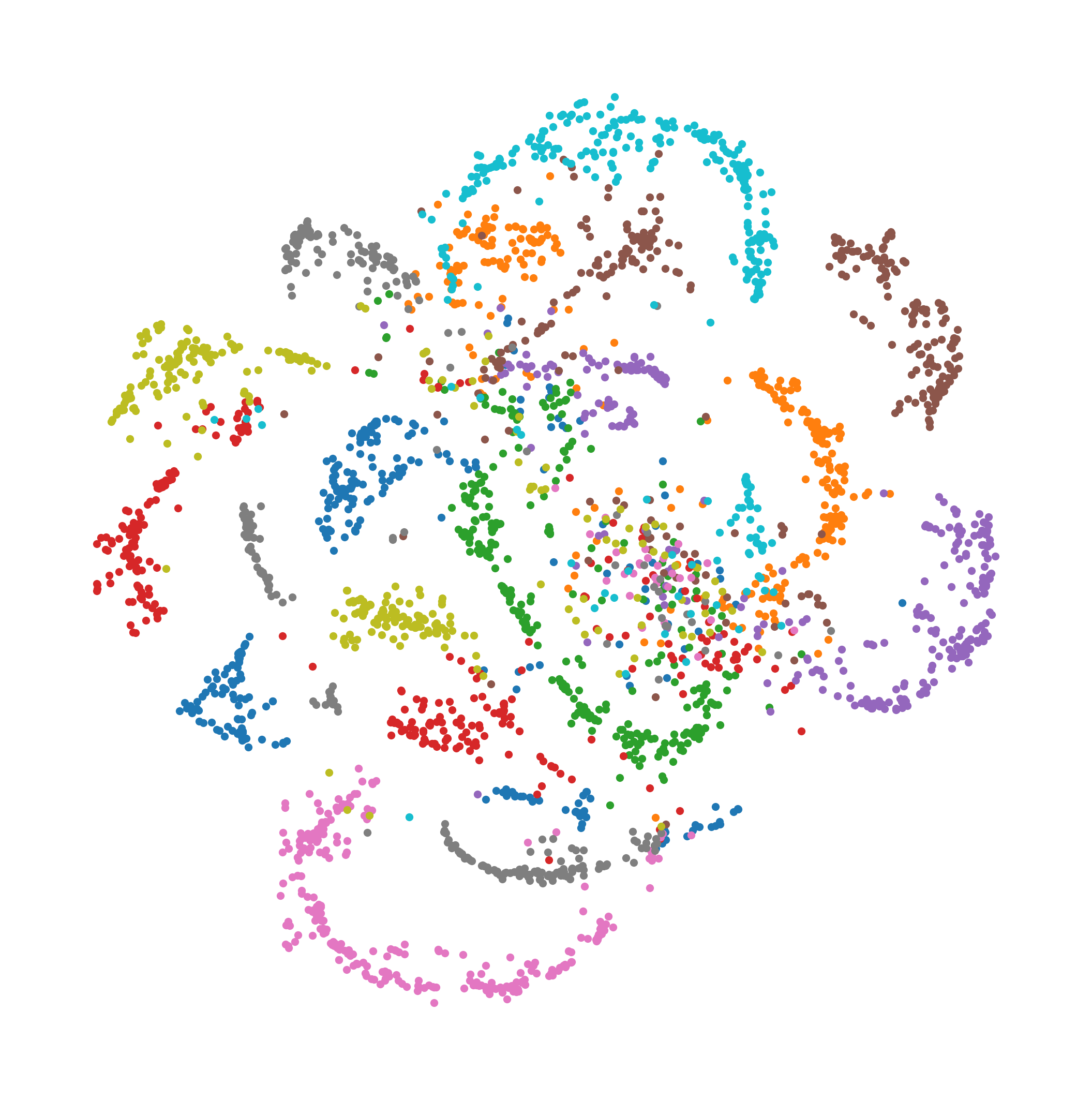}
         \caption{Pre-trained only on the LSK dataset.}
         \hspace{1.0cm}
         \vspace{-2mm}
         \label{fig:tsne_AP}
     \end{subfigure}
     \begin{subfigure}[b]{0.3\textwidth}
         \centering
         \includegraphics[width=0.65\textwidth]{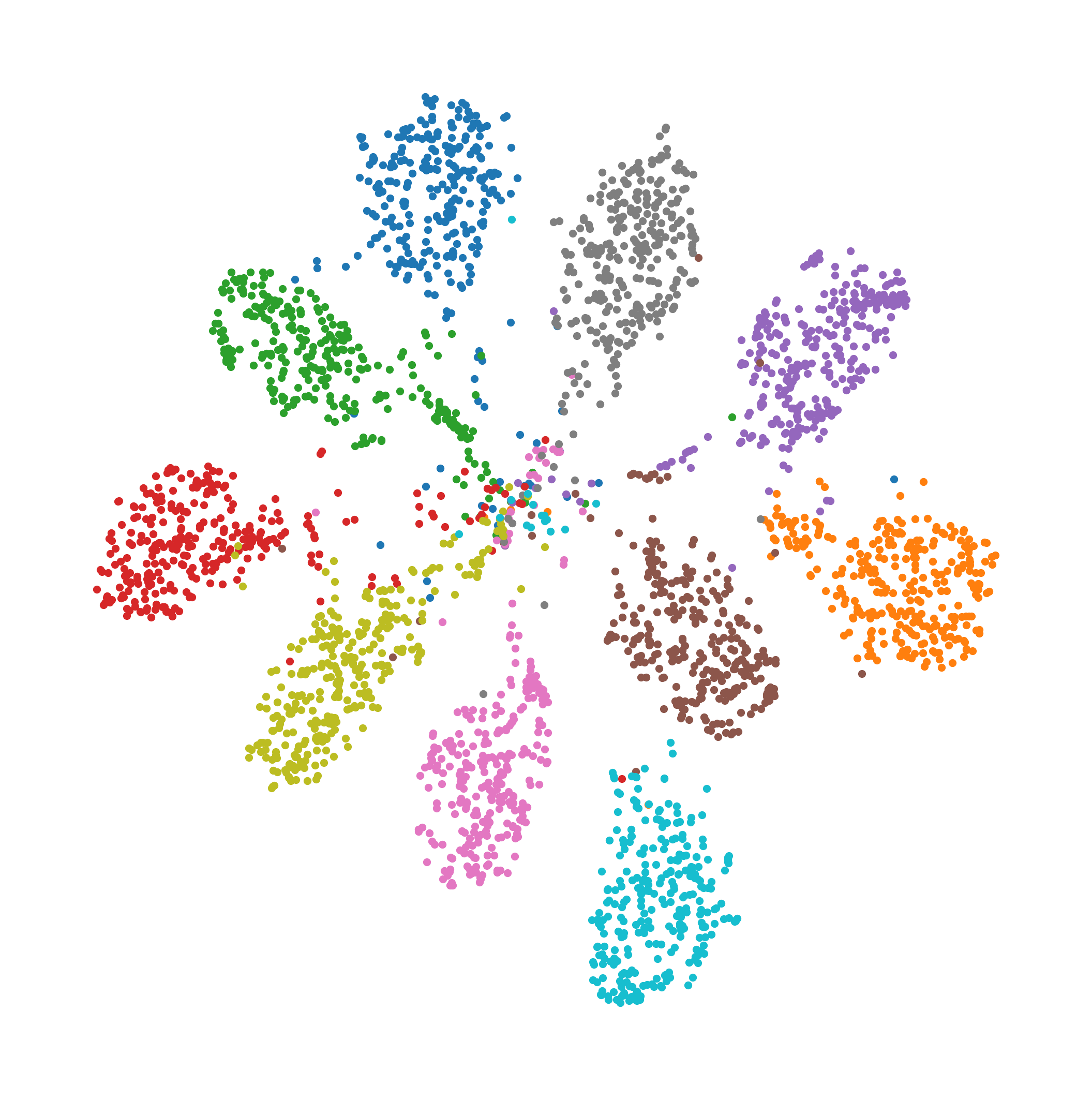}
         \caption{Pre-trained on the LSK dataset, then fine-tuned on the GSC dataset.}
         \label{fig:tsne_AP_AP}
         \vspace{-2mm}
     \end{subfigure}
     
}

\caption{t-SNE visualisation of the embedding of unseen keywords. All different colours indicate different keywords. When pre-training and fine-tuning are conducted, the model better represents user-defined keywords in the distinctive embedding space.}
\label{fig:tsne}
\vspace{-5mm}
\end{figure*}

\subsection{Evaluation Protocol}

We reserve a specific number of samples from each target keyword for enrollment. We use the centroid of the embedding features extracted from the model as the prototype for each class.
The model maps the input query signal to the embedding space and makes predictions based on the distance from the embedded features to the prototypes. We conduct experiments with 1, 5, and 10 samples for enrollment, referred to as 1-shot, 5-shot, and 10-shot enrollment, respectively.

The following metrics are employed to evaluate the performance of our system:

\newpara{Equal Error Rate (EER).}
For a particular threshold value, two types of error rates are computed to evaluate how accurately the KWS model detects user-defined keywords: FRR indicates the proportion where the target keyword is not detected by the model, and FAR indicates the proportion where non-target keyword is detected as a target keyword. The trade-off between FRR and FAR can be visualised on a DET curve by changing the threshold (See our project page). The Equal Error Rate (EER) is determined where FAR and FRR are equal.

\newpara{False Rejection Rate (FRR) at given False Alarm Rate (FAR).}
These metrics represent different operating points on the DET curve described above. In the industry, system designers are often interested in the detection rate for a fixed number of false alarms per hour (\eg~70\% detection rate at 5 false alarms). FRR at given FAR is measured to best simulate this requirement in the experiments.

\newpara{F1-score.} 
The F1-score combines the precision and recall of a classifier into a single metric by taking their harmonic mean. It is typically used to evaluate binary classification systems. 

\newpara{Accuracy.} 
Accuracy represents the ratio of the number of correct predictions to the number of tests in total in a simple 10-way classification setup. 

\subsection{Implementation Details}
\newpara{Data preprocessing.}
We extract a 40 dimensional Mel-Frequency Cepstrum Coefficient (MFCC) with a 30ms window and 10ms frame shift. To augment the training data, we add diverse noises to the input data using the MUSAN dataset~\cite{snyder2015musan} and apply various Room Impulse Response (RIR) filters. The length of audio data is set to 1 second using truncation or zero padding operation.

\newpara{Training Details.}
We select the \texttt{res15} architecture proposed in~\cite{tang2018deep} as our baseline network. The network is optimised by the Adam optimizer~\cite{kingma2014adam}. For pre-training, the batch size is set to 256 and the initial learning rate is $10^{-3}$. 
For fine-tuning, the batch size is set to 16 and the learning rate is initialised to $10^{-5}$.
During both pre-training and fine-tuning, we use a learning rate decay of 0.95 every epoch. Our framework is implemented with PyTorch~\cite{paszke2017automatic}.

\section{Results}
In this section, we analyse the effects of various objective functions, pre-training methods, and data pre-processing. All models are pre-trained on the LSK dataset and fine-tuned on the GSC dataset unless otherwise stated.

\newpara{Analysis on Objective Functions.}
The choice of objective function is paramount in learning effective representations. \Tref{tab:class_vs_metric} reports the experimental results according to the various loss functions. 

When one-stage training is applied without pre-training, the softmax and AM-Softmax show reasonable performance. The AP loss shows weak performance, since the loss requires sufficient number of classes to learn distinctive embeddings, but the limited number of classes in the GSC dataset hinders models' generalisation. On the other hand, when the pre-training and fine-tuning are performed, the model trained with AP loss in both stages shows the best performance on most metrics.
Comparing this model to the baseline model trained with the softmax loss on the only GSC dataset, the performance gap between them stands out.

\newpara{Effect of Pre-training.}
As reported in~\Tref{tab:class_vs_metric}, the pre-trained models without the fine-tuning stage do not perform well because LSK dataset's characteristics are different from that of GSC in terms of acoustics and word isolation. 
When the model is fine-tuned with the softmax loss, the performance of the model deteriorates when compared to the model trained with only the GSC dataset.
On the other hand, fine-tuning the model with metric learning-based objective functions enhances the performance of the model.
In particular, the model pre-trained and fine-tuned with AP loss outperforms the other models.

In addition, when 1-shot enrollment is conducted, the performance of our best model is even better than that of the baseline~\cite{liu2021keyword} which uses 10-shot enrollment.

\newpara{Ablation on the quantity of pre-training data.}
We perform ablations on the dataset configuration by changing (1) the number of the samples per keyword, and (2) the number of keywords while maintaining the number of samples per keyword. In both cases, the total number of samples is the same.
As shown in~\Cref{tab:samples}, reducing the number of samples or the number of classes degrades the KWS performance. The results emphasise that the number of classes is a more important factor for the model to generalise well to unseen user-defined keywords. From this observation, we further enlarge the number of classes by adding the KSK dataset to the LSK dataset. When pre-trained on both datasets, the model gains performance even when the additional dataset is composed of another language.

\newpara{Qualitative Results.}
We visualise the embeddings extracted from KWS models on the t-SNE plot~\cite{van2008visualizing} in \cref{fig:tsne}. The plot shows how well our model separates the user-defined keywords for each of the different training strategies.
As illustrated in~\cref{fig:tsne_GSC,fig:tsne_AP}, the models trained on the GSC dataset or only pre-trained on the LSK dataset cannot extract distinctive embeddings from the unseen audios. On the other hand, \cref{fig:tsne_AP_AP} shows that the proposed training strategy, where the model is pre-trained and fine-tuned sequentially, improves both inter-class compactness and intra-class separation in the embedding space.

\newpara{Effect of CER-based filtering.}
We verify the effect of the proposed CER-based filtering method in~\Tref{tab:filter}. Except for the use of data filtering, all other training details are constant. Comparing the performance, the model trained on the filtered data outperforms the model trained on the unfiltered data. Therefore, we confirm that removing the misaligned audio samples using the filtering process has a positive effect on performance.

\begin{table}[h!]
\centering
\resizebox{0.95\linewidth}{!}{
\begin{tabular}{c |ccccc} 
\specialrule{.1em}{.05em}{.05em}
Filtering & EER~$\downarrow$ & Acc~$\uparrow$ & F1-score~$\uparrow$ & FRR@FAR=2.5~$\downarrow$ & FRR@FAR=10~$\downarrow$ \\
\hline
\xmark          
                        & 3.47     & 95.67  & {\bf0.96}  & 4.83 & 1.47 \\ 
\cmark              
                        & {\bf3.24}     & {\bf95.97}  & {\bf 0.96} &  {\bf 4.20} & {\bf 1.20}     \\
\specialrule{.1em}{.05em}{.05em}
\end{tabular}}
\vspace{-3mm}
\caption{Results with and without the proposed CER-based filtering. All numbers are in percent (\%) except for F1-score.}
\vspace{-4mm}
\label{tab:filter}
\end{table}

\section{Conclusion}
\label{sec:discussion}

In this paper, we have proposed a novel metric learning-based training strategy for user-defined KWS task to represent spoken keywords in the discriminative embedding space. We collect large-scale out-of-domain keyword data, LSK dataset, to pre-train a model that can be utilised in various KWS-based downstream tasks. Furthermore, we fine-tune the model on in-domain but non-overlapping classes to improve generalisation to the target task.

We provide extensive ablations of the different factors that can affect the user-defined KWS performance. The experiments demonstrate that our best model achieves the state-of-the-art performance on the user-defined KWS task. The code, data and evaluation protocol will be released to facilitate comparison among future works.

\clearpage
\bibliographystyle{IEEEbib}
\bibliography{shortstrings,refs}

\end{document}